
\magnification=\magstep1
\baselineskip=24pt
\line{\hfill UCD-93-06}
\line{\hfill RU-93-09}
\line{\hfill FSU-SCRI-93-37}
\bigskip
\centerline{\bf The Hausdorff dimension of random walks and}
\centerline{\bf the correlation length critical exponent
in Euclidean field theory}
\bigskip
\centerline{
Joe Kiskis\footnote{$^1$}{Department of Physics,
University of California, Davis, CA 95616},
Rajamani Narayanan\footnote{$^2$}{Department of Physics,
Rutgers University, Piscataway, NJ 08855-0849},
Pavlos Vranas\footnote{$^3$}{SCRI,
Florida State University, Tallahassee, FL 32306}}
\vfill
\centerline{\bf Abstract}
\bigskip
We study the random walk representation of the
two-point function in statistical mechanics models near
the critical point. Using standard scaling arguments we
show that the critical exponent $\nu$ describing the
vanishing of the physical mass at the critical point
is equal to $\nu_\theta/ d_w$. $d_w$ is the Hausdorff
dimension of the walk. $\nu_\theta$ is the exponent describing
the vanishing of the energy per unit length of the walk at the
critical point. For the case of O(N) models, we show that
$\nu_\theta=\varphi$, where  $\varphi$ is the crossover exponent
known in the context of field theory. This implies that
the Hausdorff dimension of the walk is $\varphi/\nu$ for O(N)
models.
\bigskip
\noindent KEY WORDS: Random walks in field theory; Hausdorff dimension
of random walks; correlation length exponent.
\vfill\eject
\noindent {\bf 1. Introduction}
\bigskip
The two-point function is a quantity of central interest
in statistical mechanics models including lattice regulated quantum field
theories.
Let $G(r,t)$ be the two point function between $0$ and $r$ in a model
with one parameter $t$. Let $t=0$ be a critical point. Near $t=0$, the
two point function in three dimensions has a scaling form of the
type$^{(1)}$
$$G(r,t)={1\over r^{1+\eta}}g (rt^\nu).\eqno{(1)}$$
The scaling function $g(x)$ decays exponentially for large $x$ so that
$$m_p\sim t^\nu\eqno{(2)}$$
is the physical mass. For free field theory,
$\nu={1\over 2}$, and a departure from that
typically signifies non-trivial interactions in the underlying
field theory.

In this paper,
we present a physical understanding of the exponent $\nu$ based on
random walks.
The two-point function in
any statistical mechanics model admits a random
walk representation$^{(2)}$. This representation is
obtained by writing the two-point function as a sum of an
infinite number of terms with a one-to-one correspondence between each
term and a random walk between the two
points.

In section 2, we show that the random walk representation
leads to an expression for
 the two-point function $G( r,t)$
of the form
$$G( r,t)=\int_0^\infty dl \ S(l,t)P(l,r,t).\eqno{(3)}$$
$S(l,t)$ is the energy factor, and $P(l,r,t)$ is the entropy factor.
$P(l,r,t)$ is the probability density of walks of fixed
length $l$:
$$P(l,r,t)\ge 0,\ \ \ \forall l,r,t;\ \ \ \ {\rm and}\ \ \ \
\int d^3r\  P(l,r,t)=1, \ \ \ \ \forall l,t.\eqno{(4)}$$
Just as the two-point function has a scaling form near $t=0$,
we assume that both $S(l,t)$ and $P(l,r,t)$ have scaling forms
near $t=0$.
We will be interested in the Hausdorff dimension $d_w$ of the
walk near $t=0$.

The mean distance $R(l)$ of a walk of length $l$
 is proportional to the square root of
the second moment of the probability density  with respect to $r$.
For $t \rightarrow 0$, the probability distribution of
walks will be spread out, and $R(l)$ will diverge  for large $l$
with a leading behavior of the type
$$R(l)\sim l^{\nu_w}.\eqno{(5)}$$
$d_w=1/\nu_w$ is the Hausdorff dimension of the walk.

The energy factor $S(l,t)$
 is the weighted sum of all walks of length $l$. Therefore,
$\theta(t) \equiv -{1\over l}\ln S(l,t)$ for
$l\rightarrow\infty$ is an appropriate
definition of the energy per unit length
of the walk. Near $t=0$
walks of all lengths become equally important which implies
that $\theta(t)$ vanishes as $t\rightarrow 0$. We assume the leading
behavior
$$\theta(t)\sim t^{\nu_\theta}.\eqno{(6)}$$
$\nu_\theta$ is another quantity of interest to us.
Both (5) and (6) are similar in spirit to (2)
and  follow from the scaling form we assume for
$P(l,r,t)$ and $S(l,t)$ just as (2) follows from (1).

The scaling forms for $S(l,t)$ and $P(l,r,t)$
when combined in (3) must be consistent with (1).
Using this we show in section 2
that
$$\nu=\nu_w\nu_\theta.\eqno{(7)}$$
The content of (7) is that the nonanalyticity in (2) near $t=0$
has two sources:
a) a nonanalyticity in $\theta (t)$, which controls the length $l$, and
b) a nonanalyticity in the relation $R(l)$ between the mean distance and the
length.
 If the underlying theory in the vicinity of the
critical point is free, then $\nu_\theta=1$ and $\nu_w={1\over 2}$.

We consider O(N) models in section 3 and show that for these models,
$$\nu_\theta=\varphi,\eqno{(8)}$$
where $\varphi$ is a cross-over exponent$^{(3)}$.
Substitution of (8) in (7) gives
$$d_w={1\over \nu_w}={\varphi\over\nu}.\eqno{(9)}$$
\bigskip
\noindent {\bf 2. Derivation of the scaling law $\nu=\nu_w\nu_\theta$}
\bigskip
The two-point function $G( r,t)$
 can always be cast in the form$^{(2)}$,
$$G(r,t)=\sum_{w:0\rightarrow r}\Omega(w,t).\eqno{(10)}$$
The sum is over all random walks $w$ connecting $0$ and $r$, and
(10) is called the random walk representation of the two-point
function. $\Omega(w,t)$ depends upon the model
and is always positive. By grouping together all
walks of length $l$, (10) can be rewritten as
\footnote{$^4$}{In the
following equation, the integral is to be understood as a limit of a
sum.},
$$G(r,t)=\int_0^\infty dl\ \tilde\Omega(l,r,t).\eqno{(11)}$$
Define
$$S(l,t)=\int d^3r\ \tilde\Omega(l,r,t),\eqno{(12)}$$
and
$$P(l,r,t)={\tilde\Omega(l,r,t)\over S(l,t)}.\eqno{(13)}$$
Substitution of (13) in (11) gives (3) with (13) satisfying the
property (4).

In a manner similar to (1), both $S(l,t)$ and $P(l,r,t)$ are expected
to have scaling forms. For $S(l,t)$, we assume the following general form:
$$S(l,t)={1\over l^{\eta_p}}s(lt^{\nu_\theta})\eqno{(14)}$$
with an exponential decay for $s$ at large argument.
This is identical to the form for $G(r,t)$ in (1).
{}From (14), it is clear that the energy per unit length $\theta(t)$
is given by (6).

Next, we have to write down the scaling form for $P(l,r,t)$.
Here we
have to keep in mind that it has to be in concordance with (4), and
with the combination of (1), (3), and (14).
This results in the
following general form for $P(l,r,t)$:
$$P(l,r,t)={1\over r^3}p(rl^{-\nu_w},rt^\nu). \eqno{(15)}$$
The pre-factor $1/r^3$ is in agreement with (4).
The leading behavior of the mean distance of the walk for large
$l$ is given by (5).
The combination $rt^\nu$ in (15) is
in accordance with (1) and (3).

Further, consistency of (14), (15), and
(3) with (1)
gives the scaling law
(7). It also results in a
scaling relation for the anomalous dimension $\eta_p$,
$$\eta_p=1-(2-\eta)\nu_w.\eqno{(16)}$$

The scaling relations (7) and (16) involve three exponents defined in the
walk picture: $\nu_w$, $\nu_\theta$ and $\eta_p$. In the next section,
we focus on the O(N) models to obtain (8). The equations (8),
(7), and (16) are
three relations for those three exponents.
\vfill\eject
\noindent{\bf 3. O(N) models and $\nu_{\theta}$ }
\bigskip
At fixed spatial cutoff $a=1/\Lambda$, the O(N) symmetric Lagrangian
is
$${\cal L}(\vec\phi)={1\over 2}(\partial\vec\phi\cdot\partial\vec\phi)+{1\over
2}(m_c^2+t)(\vec\phi\cdot\vec\phi)+{\lambda_c\over 4!}
(\vec\phi\cdot\vec\phi)^2.\eqno{(17)}$$
$m_c$ and $\lambda_c$ are chosen so that $t=0$ is the critical point.

The first step in deriving the random walk representation for the two
point function is to write the interaction term as
$$e^{-{\lambda_c\over 4!}(\vec\phi\cdot\vec\phi)^2}=
{1\over \sqrt{2\pi}}\int d\sigma \ e^{-{1\over
2}\sigma^2+b\sigma(\vec\phi\cdot\vec\phi)},\eqno{(18)}$$
with
$$b=i\sqrt{2\lambda_c\over 4!}.\eqno{(19)}$$
The partition function becomes
$$\eqalign{Z(t)&=\int [d\vec\phi][d\sigma]\ e^{-{1\over 2}\int d^3x\
\Bigl[\sigma^2+\vec\phi\cdot(H+t)\vec\phi\Bigr]}\cr
&=\int[d\sigma]
 e^{-{N\over 2}{\rm Tr}\ln(H+t)-{1\over 2}\int d^3x\ \sigma^2},\cr}
\eqno{(20)}$$
with
$$H=-\partial^2+m_c^2-2b\sigma.\eqno{(21)}$$
The two-point function,
$$\eqalign{G(r,t)\delta^{ij}&=<\phi^i(0)\phi^j(r)>,\cr
&=\int_0^\infty dl\ e^{-tl}
{1\over Z(t)}
\int[d\sigma] e^{-{N\over 2}{\rm Tr}\ln(H+t)-{1\over 2}\int d^3x\
\sigma^2}\Psi(l,r).\cr}\eqno{(22)}$$
In (22), $\Psi(l,r)$, is the solution to
$$-{\partial\over\partial l}\Psi(l,r)=H\Psi(l,r),\eqno{(23)}$$
with the initial condition,
$$\Psi(0,r)=\delta^{3}(r).\eqno{(24)}$$
$\Psi(l,r)$ is the propagation kernel for the Hamiltonian H.
It has a standard path integral representation in which the
sum is over all paths from
the origin to $r$ in proper time $l$.
The length $L$ of these paths
is proportional to
$l$ and inversely proportional to the spatial cutoff $a$.

Therefore the integrand in (22) is indeed $\tilde\Omega(l,r,t)$
as defined in (11).\footnote{$^5$}{To make the explicit
connection to (10) one just
has to write down the path integral sum of $\Psi(l,r)$.}
Hence $S(l,t)$, defined in (12), can be written as
$$S(l,t)=e^{-tl}S_I(l,t)\eqno{(25)}$$
where
$$S_I(l,t)=
{1\over Z(t)}
\int[d\sigma] e^{-{N\over 2}{\rm Tr}\ln(H+t)-{1\over 2}\int d^3x\
\sigma^2}\Bigl[\int d^3r\Psi(l,r)\Bigr].\eqno{(26)}$$
$S_I(l,t)$ incorporates the interaction of the
walk in $\Psi(l,r)$ with the background loops in ${\rm Tr}\ln(H+t)$.
If there is no interaction term in (17)
($\lambda_c=0$), then $S_I(l,t)=1$,
$\eta_p=0$, and $\nu_\theta=1$.

The key issue is to understand
the behavior of $S_I(l,t)$ when $N\ge 1$. In
particular,
the energy per unit length of the walk is
$$\theta(t)=t-\lim_{l\rightarrow\infty}{\ln (S_I(l,t))\over
l}.\eqno{(27)}$$
The second term on the right hand side of (27) is the contribution
from the interaction of the walk with the background loops.
We would like to know its behavior near
$t=0$.
We will now show that for small t,
$$\lim_{l\rightarrow\infty} {\ln (S_I(l,t))\over l}=
t-ct^\varphi+{\rm\ higher\ powers\ in\ }t.\eqno{(28)}$$
This will result in (6) and (8).

Toward this end, we softly break the O(N) symmetry in (17) to an
O(K)$\times$O(N-K) symmetry by introducing a different mass term
for the K-component $\vec\phi_1$ field and the (N-K)-component
 $\vec\phi_2$ field\footnote{$^6$}{The first K components of
$\vec\phi$ form the
vector $\vec\phi_1$ and the last (N-K) components of $\vec\phi$ form
the vector $\vec\phi_2$}:
$$(m_c^2+t)(\vec\phi\cdot\vec\phi)\rightarrow
(m_c^2+t)(\vec\phi_1\cdot\vec\phi_1) +
(m_c^2+t')(\vec\phi_2\cdot\vec\phi_2).\eqno{(29)}$$
 $t=t'=0$,
where the O(N) symmetry is broken, is the critical point of interest to
us. But for the model with
the asymmetric term (29), there are two critical lines in the
$(t,t')$ plane$^{(3)}$. One critical line corresponds to the breaking of
the O(K) symmetry, and the other line corresponds to the breaking of
the O(N-K) symmetry.  These lines meet at the bicritical point
$t=t'=0$ where the O(N) symmetry is broken.

It is useful$^{(3)}$ to define the effective thermal parameter $T$ as
$$T={Kt+(N-K)t'\over N},\eqno{(30)}$$
and the anisotropy $g$ as
$$g={t'-t\over N}.\eqno{(31)}$$
The critical line where the O(K) symmetry is broken is given by$^{(3)}$
$$T=(\alpha g)^{1\over\varphi}+{\rm\ higher\ powers\ in\ }g.\eqno{(32)}$$
Near $t=t'=0$, (32) can be rewritten as
$$t=t'-c{t'}^{\varphi}+{\rm\ higher\ powers\ in\ }t',\eqno{(33)}$$
with $c=N/\alpha$.
$\varphi$ is called the crossover exponent,
and the $\epsilon$-expansion gives$^{(3)}$
$$\varphi =1+{N\over 2(N+8)}\epsilon+{N^3+24N^2+68N\over
4(N+8)^3}\epsilon^2+O(\epsilon^3)>1.\eqno{(34)}$$
In this, $\epsilon=4-d$, and $d$ is the Euclidean dimension in which the
model is defined.
 Let us consider the two-point function of the
$\vec\phi_1$ field
$$G(r,t,t')\delta^{ij}=<\phi^i_1(0)\phi^j_1(r)>.\eqno{(35)}$$
As before, we can write down the random walk representation of this
two-point function. Instead of (25), we will now have
$$S(l,t,t')=e^{-tl}S_I(l,t,t'),\eqno{(36)}$$
where
$$S_I(l,t,t')=
{1\over Z(t,t')}
\int[d\sigma] e^{-{K\over 2}{\rm Tr}\ln(H+t)
-{N-K\over 2}{\rm Tr}\ln(H+t')-{1\over 2}\int d^3x\
\sigma^2}\Bigl[\int d^3r\Psi(l,r)\Bigr],\eqno{(37)}$$
and
$$Z(t,t')
=\int[d\sigma] e^{-{K\over 2}{\rm Tr}\ln(H+t)-{N-K\over 2}{\rm Tr}\ln(H+t')
-{1\over 2}\int d^3x\ \sigma^2}.
\eqno{(38)}$$
The energy per unit length of the walk is
$$\theta(T,g)=t-f(t,t'),\eqno{(39)}$$
with
$$f(t,t')=\lim_{l\rightarrow\infty} {\ln(S_I(l,t,t'))\over
l}.\eqno{(40)}$$
When $t=t'$, (37) goes to (26), (38) goes to (20), and
(39) goes to (27).

To proceed further, we have to look at the scaling form
of $\theta(T,g)$ near the critical line (32).
{}From the discussion in Ref. 3, it follows that $\theta$ has
the form
$$\theta(T,g)=T^{\nu_\theta}h_K(g/T^{\varphi}).\eqno{(41)}$$
$h_K(x)$ is regular at $x=0$, and the leading behavior in (41)
is compatible with (6) when $g=0$. (39) and (41) have to be
consistent. For example, the calculations of
$${\partial\theta(T,g)\over\partial g}|_{{}^{g=0}_{K=0}}\eqno{(42)}$$
from (39) and (41) must agree. This gives
$$-N=t^{\nu_\theta-\varphi}h_0^\prime(0).\eqno{(43)}$$
$h_K^\prime(x)$ is the derivative of $h_K(x)$ with respect to $x$.
The left hand side of (43) is obtained by evaluating (42) using (39).
We have used (30) and (31) and the fact that $f(t,t')$ in (39)
does not depend on $t$ when $K=0$ (see (37) and (40)). The right hand
side of (43) is obtained by evaluating (42) using (41). $h_0(x)$ is
regular at $x=0$ and $h_0^\prime(0)$ is a constant. (43) is valid
for a range of $t$ near $0$. This means that $h_0^\prime(0)=-N$ and
more significantly,
the relation of interest (c.f.(8)), $\nu_\theta=\varphi$.
\bigskip
\noindent{\bf 4. Conclusions}
\bigskip
In this paper we analysed the singular behavior of the random walk
representation of the two-point function. The energy per unit length
of the walk $\theta(t)$ is found to have a non-trivial dependence on
the bare parameter $t$ (c.f.(6)).
This is attributed to the fact that the walk is taking place in the
presence of background loops.
For O(N) models, the exponent
$\nu_\theta$ characterizing the nonanalytic behavior of $\theta(t)$
is shown to be same as another exponent already known
in the context of field theory, namely the
crossover exponent $\varphi$ (c.f.(8)).
This connection enables us to derive a relation for the Hausdorff
dimension of the walk in terms of standard exponents in field theory
(c.f.(9)).
\bigskip
\noindent {\bf Acknowledgements}

This research was supported in part by the DOE under
grant \# DE-FG03-91ER40674 (JK),
grant \# DE-FG05-90ER40559 (RN)
and under grants \# DE-FG05-85ER250000 \&
\# DE-FG05-92ER40742 (PV).
\bigskip
\noindent {\bf References}

\item {1.} D.~J.~Amit, {\sl Field Theory, the Renormalization Group, and
Critical Phenomena} {\bf World Scientific} (1984) I-1-2.

\item {2.} J.~Glimm and A.~Jaffe, {\sl Quantum Physics } {\bf
Springer-Verlag} (1987) chapter 21;
C.~Itzykson and J.~Drouffe, {\sl Statistical
Field Theory} {\bf Cambridge University Press} (1989) 1.2.2.

\item {3.} D.~J.~Amit, {\sl Field Theory, the Renormalization Group, and
Critical Phenomena} {\bf World Scientific} (1984) II-5-3.

\item {4.} C.~Itzykson and J.~Drouffe, {\sl Statistical
Field Theory} {\bf Cambridge University Press} (1989) 1.2.3.

\end